\documentclass[fp,onecolumn]{jpsj3}
\usepackage{txfonts}
\usepackage{bm}
\usepackage{color}
\usepackage{ulem}
\usepackage[dvipdfmx]{graphicx}
\usepackage{dcolumn}
\usepackage{mathrsfs}
\usepackage{ulem}
\usepackage{xcolor}
\title{Magnetic Instability of Pr$_3$Ru$_4$Sn$_{13}$}

\author{Takanori Taniguchi$^1$\thanks{takanori.taniguchi.d3@tohoku.ac.jp}, Shinnosuke Kitayama$^{1,2}$, Hirotaka Okabe$^1$, Jumpei G. Nakamura$^3$, Akihiro Koda$^3$, Motoyuki Ishikado$^4$, and Masaki Fujita$^1$}
\inst{$^1$Institute for Materials Research, Tohoku University, Sendai 980-8577, Japan \\
$^2$Department of Physics, Tohoku University, Sendai 980-8578, Japan \\
$^3$Institute of Materials Structure Science, High-Energy Accelerator Research Organization, Tsukuba 305-0801, Japan \\
$^4$Neutron Science and Technology Center, Comprehensive Research Organization for Science and Society, Tokai 319-1106, Japan} 

\abst{ 
$\;$We report on the quantum criticality of Pr$_3$Ru$_4$Sn$_{13}$ revealed by our new material research. Pr$_3$Ru$_4$Sn$_{13}$ has been synthesized by flux growth and characterized by single X-ray, powder X-ray, and powder neutron diffraction measurements. The compound adopts a Yb$_3$Rh$_4$Sn$_{13}$-type structure with a cubic Pm$\bar{3}$n. From the magnetization at 1 T, the effective magnetic moment was estimated to be 3.58 $\mu _B$ per Pr$^{3+}$, suggesting that the magnetism is mainly contributed by Pr$^{3+}$ ions. The specific heat and magnetization show an anomaly at $T_{N} = 7.5$ ~ K owing to the phase transition. The muon spin rotation and relaxation ($\mu$SR) time spectra exhibit clear oscillations below $T_N$. This suggests that the phase is magnetically ordered. The volume fraction of the magnetic phase estimated from the initial asymmetry is around ten percent. In addition, spin fluctuations were observed at low temperatures. These results provide microscopic evidence that the material is closest to the antiferromagnetically quantum critical point with a partial order among Pr$_3$$T_4$Sn$_{13}$ ($T= $ Co, Ru, Rh).}

\begin{document}
\maketitle

\section{Introduction}

$c-f$ hybridization provides various physical properties such as long-range order, Fermi liquids, non-Fermi liquids, and unconventional superconductivity~\cite{Stewart_1984}.
This is due to the competition within the Ruderman$\textendash$Kittel$\textendash$Kasuya$\textendash$Yosida (RKKY) interaction, which stabilizes the magnetic moment, and the Kondo effect, which vanishes the magnetic moment.
CeRhIn$_5$~\cite{Hegger_2000} and CeCoIn$_5$~\cite{Petrovic_2001} show antiferromagnetism and superconductivity because the magnitudes of the RKKY interaction and Kondo effect are changed by the chemical pressure effect~\cite{Shimozawa_2016}.
In particular, CeCoIn$_5$ at the quantum critical point (QCP) has been reported to have unique physical properties, such as a Fulde$\textendash$Ferrell$\textendash$Larkin$\textendash$Ovchinnikov (FFLO) state~\cite{Kenzelmann_2008, Young_2007}. The region around the QCP is rich in anomalous physical properties.

The search for exotic quantum criticality in the caged Pr1-2-20 family has recently been active.
PrTi$_2$Al$_{20}$ shows $O_{20}$-type ferroquadrupole order at ambient pressure~\cite{Sakai_2011, Taniguchi_2016, Taniguchi_2019}.
When pressure is applied, the ferroquadrupole order is suppressed, and heavy-fermion superconductivity appears~\cite{Matsubayashi_2012, Matsubayashi_2014}.
PrV$_2$Al$_{20}$~\cite{Tsujimoto_2014} exhibits superconductivity at ambient pressure.
Since the smaller lattice constant of PrV$_2$Al$_{20}$ than that of PrTi$_2$Al$_{20}$, PrV$_2$Al$_{20}$ exists near the QCP because of the chemical pressure effect.


\begin{figure}[h]
\begin{center}
\includegraphics[width=9cm,clip]{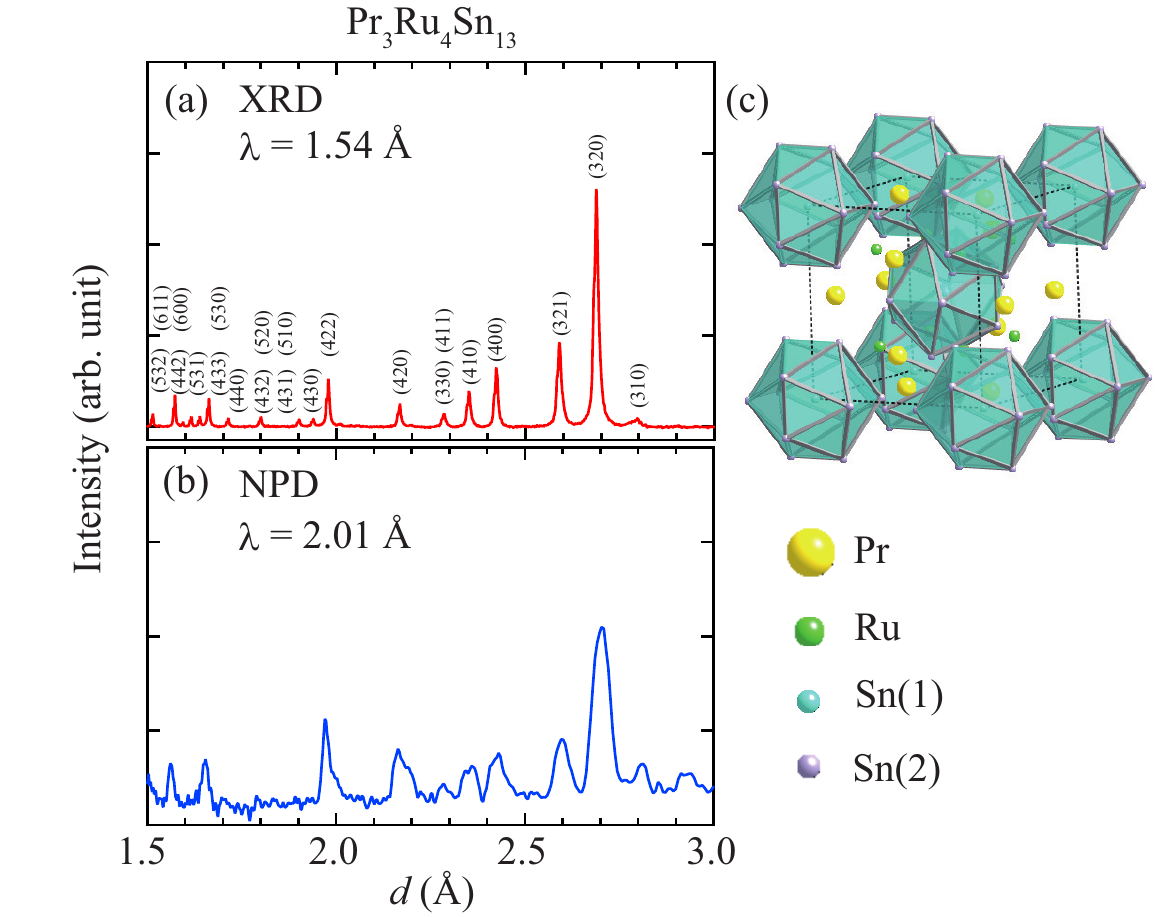}
\caption{\label{fig:crystal_structure} (Color online) (a) X-ray powder diffraction (XRD) and (b) neutron powder diffraction (NPD) patterns of Pr$_3$Ru$_4$Sn$_{13}$. (c) Crystal structure of Pr$_3$Ru$_4$Sn$_{13}$.
}
\end{center} 
\end{figure}

\begin{table}[h]
\caption{Crystal data and refinement detail of Pr$_3$Ru$_4$Sn$_{13}$.}
\label{t1}
\begin{center}
\begin{tabular}{cc}
\hline 
Crystal formula & Pr$_{3}$Ru$_{4}$Sn$_{13}$\tabularnewline
Crystal system, space group & cubic, Pm$\bar{3}$n\tabularnewline
$a=b=c$ ($\text{\AA}$) & 9.7157(3)\tabularnewline
$Z$ & 2\tabularnewline
Radiation type & Mo-K$\alpha$\tabularnewline
$R_{int}$ & 0.0382\tabularnewline
$R$ indiexes {[}$I>2\sigma${]} & $R_{1}=2.81$\%, $wR_{2}=7.40$\%\tabularnewline
$R$ indiexes {[}all data{]} & $R_{1}=2.89$\%, $wR_{2}=7.43$\%\tabularnewline
\hline 
\end{tabular}
\end{center}
\end{table}

\begin{table}[h]
\caption{Pr$_3$Ru$_4$Sn$_{13}$: atomic positions and structure information at room temperature.}
\label{t1}
\begin{center}
\begin{tabular}{|c|c|c|c|c|}
\hline 
site & Wyckoff position & x & y & z\tabularnewline
\hline 
Pr & 6c & 1/2 & 1/4 & 0\tabularnewline
\hline 
Ru & 8e & 1/4 & 1/4 & 1/4\tabularnewline
\hline 
Sn (1) & 2a & 0 & 0 & 0\tabularnewline
\hline 
Sn (2) & 24k & 0 & 0.15456(5) & 0.30325(4)\tabularnewline
\hline 
\end{tabular}
\end{center}
\end{table}

Exotic quantum criticality has also been reported for the Ce3-4-13 family of skutterudite-related materials.
The quantum criticality of Ce$_3$Co$_4$Sn$_{13}$ coexisting with the CDW has been reported.
Ce$_3$Co$_4$Sn$_{13}$ has an effective magnetic moment of 2.56 $\mu _B$, which is very close to the theoretical value of 2.54 $\mu _B$, indicating that Ce$^{3+}$ contributes to the magnetism~\cite{Thomas_2006}.
The Kondo effect was observed in electrical resistivity measurements, suggesting that $f$ and conduction electrons are strongly correlated~\cite{Thomas_2006,Slebarski_2018}.
Furthermore, Ce$_3$Co$_4$Sn$_{13}$ exhibits no magnetic ordering above 0.35 K and a large specific heat of $\sim$4 J/mol K$^2$ at zero fields, making it a quantum critical material~\cite{Thomas_2006,Cornelius_2006}.
In neutron scattering experiments, antiferromagnetic correlations develop at zero magnetic field; therefore,  quantum criticality originates from antiferromagnetic spin fluctuations~\cite{Christianson_2008}.
Moreover, a CDW transition has been observed at 150 K from NMR~\cite{Lue_2012}. The XRD at the Sn site, in which the Wyckoff position is 24k, has been reported to be distorted by the CDW~\cite{Slebarski_2012}.
This transition is caused by the correlation between the $d$ electrons of Co and the $p$ electrons of Sn.
This supports the DFT calculations~\cite{Slebarski_2015,Zhong_2009} and XAS~\cite{Slebarski_2012} results, which show  a large $d-p$ interaction.

To find the new material with novel quantum criticality, we focus on the Pr3-4-13 family, which has the same crystal structure as the Ce3-4-13 family. Here, we report the single-crystal synthesis of Pr$_3$Ru$_4$Sn$_{13}$, characterized by single X-ray, powder X-ray, and powder neutron diffraction measurements, and its physical properties by magnetization, specific heat, and muon spin rotation and relaxation ($\mu$SR) measurements.

\section{Sample Preparation and Experimental Details}
Pr$_3$Ru$_4$Sn$_{13}$ single crystals were synthesized using the self-flux method.
Ingots of Pr (99.9\% purity), Ru powder (99.95\% purity), and Sn shot (99.99\% purity) were weighed and placed in  alumina crucibles in a ratio of 1:1:20 (Pr:Ru:Sn).
The samples, weighing approximately 4 g, were covered with quartz wool and encapsulated in evacuated quartz ampoules, followed by heating to 1273 K for 5 h.
Subsequently, the ampoules were cooled to 1123 K at a rate of 75 K/h, and finally cooled to 573 K at a rate of 33 K/h. At this temperature, the ampules were removed from the furnace and  excess Sn was removed by centrifugation. The chemical composition was determined to be 3:4:13 by EDX.
The phase purity of Pr$_3$Ru$_4$Sn$_{13}$ was examined using X-ray and neutron powder diffraction (XRD and NPD) measurements. In the XRD measurements, Cu-K$\alpha$ radiation with a wavelength ($\lambda$) of 1.54 \AA $ $ was used to collect the diffraction profiles using MiniFlex (RIGAKU). In addition, neutron diffraction measurements were conducted using Ge(311) reflection ($\lambda = 2.01$ \AA) on an AKANE diffractometer at Japan Research Reactor No. 3 (JRR-3), Japan.
Single-crystal X-ray diffraction data were collected using an XtaLab mini II (RIGAKU) with Mo-K$\alpha$ radiation.
Magnetic susceptibility and specific heat were measured using a Quantum Design physical property measurement system (PPMS) and a SQUID magnetometer (MPMS).
The $\mu$SR measurements were performed with a $^3$He cryostat using an S1 (ARTEMIS) $\mu$SR spectrometer at the Materials and Life Science Experimental Facility (MLF) in J-PARC, Japan.
The polarized muon beam was  stopped in the sample placed at the center of the longitudinal coils.
The time evolution of the muon spin polarization between the forward and backward counters $N$(0$^\circ$, $t$) and $N$(180$^\circ$, $t$) is placed at 0$^\circ$ and 180$^\circ$ from the beam to detect positrons. The angular distribution of decay positrons is expressed by Eq. (1); $A(t)$ is obtained by taking the time-differential ratio of the spectra:
\begin{equation}
A\left(t\right) = \frac{\left(N\left(0^{\circ},\,t\right)-N\left(180^{\circ},\,t\right)\right)}{\left(N\left(0^{\circ},\,t\right)+N\left(180^{\circ},\,t\right)\right)},
\end{equation}
after background subtraction.

\section{Experimental Results and Discussion}
\subsection{Crystal structure}

Crystal data and refinement details from single-crystal X-ray diffraction experiments on Pr$_3$Ru$_4$Sn$_{13}$ are shown in Table I. Using Olex2~\cite{Dolomanov_2009}, the structure was solved using the SHELXS~\cite{Sheldrick_2008} structure solution program using direct methods and refined with the SHELXL~\cite{Sheldrick_2015} refinement package using least Squares minimization.
The crystal structure has a cubic space group of Pm$\bar{3}$n (No. 223). The lattice constants were 9.7157(3) \AA~and Z = 2.
The atomic positions and structural information are summarized in Table II.
Figures 1(a) and 1(b) show the XRD and NPD patterns, respectively.
All the peaks can be assigned to the parameters listed in Tables I and II.
The crystal structure was solved via Rietveld analysis using the FULLPROF package~\cite{Rodriguez_1993}.
These results are consistent with $R_3$$T_4$Sn$_{13}$ ($R = $ La, Ce, $T= $ Co, Ru, Rh)~\cite{Thomas_2006,Zhong_2009,Slebarski_2012,Gamza_2008,Mishra_2011,Nair_2018}.
From these results, we conclude that the crystal structure of Pr$_3$Ru$_4$Sn$_{13}$ is as shown in Fig. 1 (c).

\begin{figure}
\begin{center}
\includegraphics[width=9cm,clip]{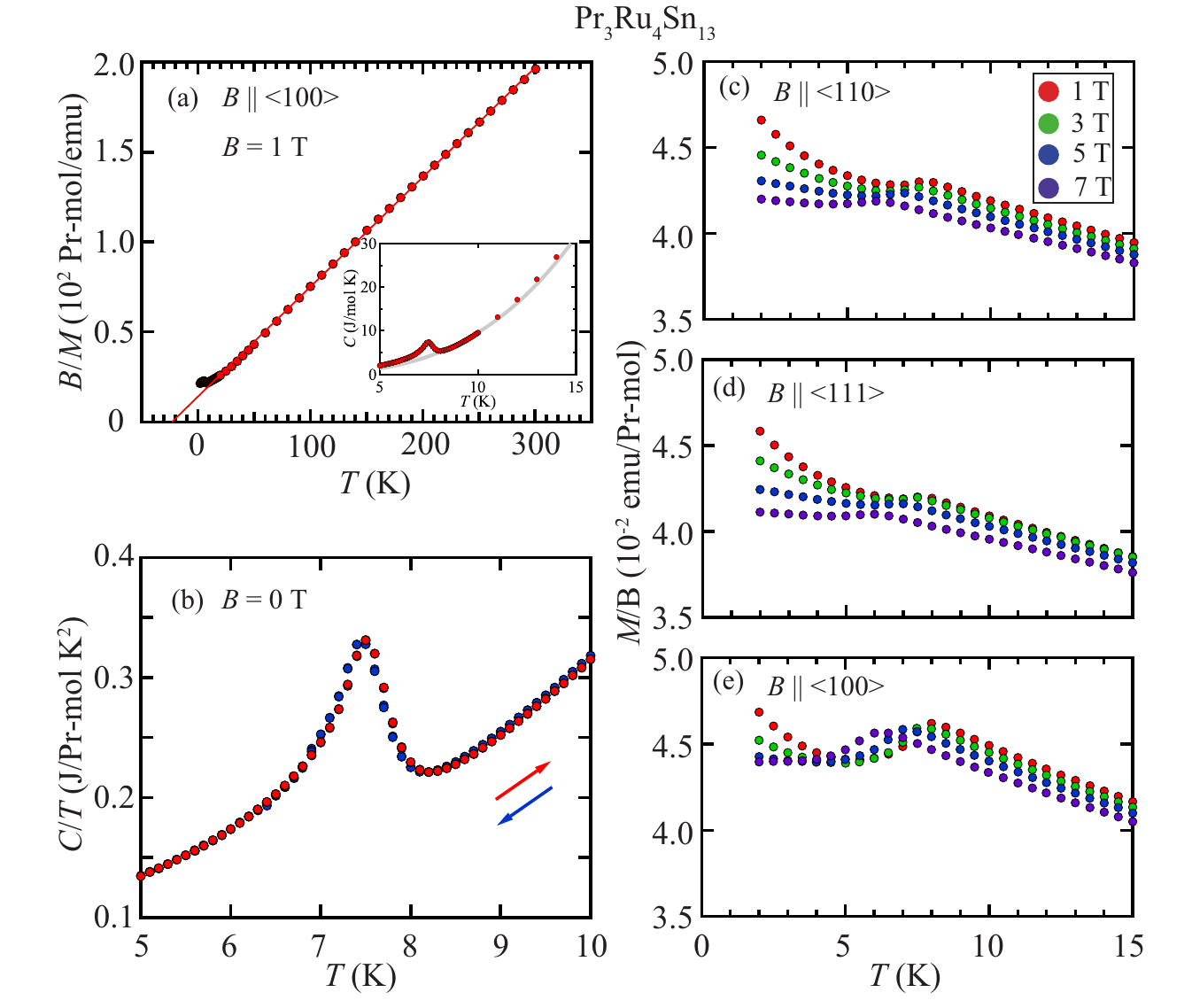}
\caption{\label{fig:crystal_structure} (Color online) (a) Temperature-dependent inverse magnetic susceptibility $B/M$ of Pr$_3$Ru$_4$Sn$_{13}$ at 1 T for $B \parallel \left\langle 100 \right\rangle$. The solid line on $B/M$ represents a fitted curve (see text). The inset shows the temperature dependence of specific heat. The solid gray line is the fitting curve (see text). (b) The results for the temperature-increasing (red) and -decreasing (blue) sweeps near the phase transition at zero-field, showing no hysteresis in the specific heat as $C/T$ process. Temperature dependence of the magnetic susceptibility $M/B$ under several magnetic fields parallel to the (c) $\left\langle 110 \right\rangle$, (d) $\left\langle 111 \right\rangle$, and (e) $\left\langle 100 \right\rangle$ axes.
}
\end{center} 
\end{figure}

\subsection{General behavior of magnetization and specific heat}

Figure 2(a) shows the temperature dependence of $B/M$ at 1 T for $B \parallel \left\langle 100 \right\rangle$, which can be described by the Curie$\textendash$Weiss rule at temperatures above 50 K.
The estimated effective magnetic moment is 3.62 $\mu _B$/Pr, which is close to the theoretical value of 3.58 $\mu _B$/Pr estimated from free Pr$^{3+}$.
This suggests that the magnetic ions Pr$^{3+}$ are localized at high temperatures and that Ru and Sn do not contribute to magnetism.
The inset shows the temperature dependence of specific heat. 
The value of specific heat was fitted using the fitting function $C=\gamma T+\alpha T^3$ between 9 K and 17 K to estimate an electronic specific heat coefficient of 40 mJ/mol K$^2$.
The 4$f$ electrons in Pr$_3$Ru$_4$Sn$_{13}$ are strongly correlated with the conduction electrons, since this value is larger than that of the normal metal.
Figure 2(b) shows the specific heat process of Pr$_3$Ru$_4$Sn$_{13}$ under zero magnetic field, where a $\lambda$-type anomaly exists near $T_N = 7.5$ K, suggesting a second-order transition.
In addition, no clear hysteresis was observed in the temperature increase and decrease scans, which agrees with the second-order transition.
In the Pr$^{3+}$,  since there is no degeneracy above 4, the minimum amount of entropy is $R$ln2 ($R$: gas constant) near the transition temperature.
However, the magnitude of the jump in specific heat is an order of magnitude smaller than that expected from $R$ln2.
There are two possible reasons for this: (i) a large electronic specific heat coefficient due to Kondo-lattice-based heavy fermions and (ii) the non-full volume fraction of the phase (partial order).
Since a large fluctuation provides a large $\gamma$ and suppresses the moment size, the fluctuation decreases the amount of entropy.
This situation has been reported for the Ce compounds CeRh$_6$Ge$_4$~\cite{Matsuoka_2015}, CeRu$_2$Ge$_2$~\cite{Bohm_1988}, CeRu$_2$Al$_2$B~\cite{Baumbach_2012}, CePd$_2$P$_2$~\cite{Tran_2014}, and CeRuPO~\cite{Krellner_2007}.
Next, to obtain the value of the magnetic specific heat $C_{mag}$ experimentally, the phonon contribution obtained from the fitting function in the inset of Fig. 2(a) was subtracted from the experimental results.
Because integration of the value of $C_{mag}$ between 6 K and 9 K yields a value of $\int dT\left(\frac{C_{mag}}{T}\right)/R\ln2\sim0.1$, the volume fraction of partial order is around ten percent.
Figures 2(c)–2(e) show the temperature dependence of $M/B$ of Pr$_3$Ru$_4$Sn$_{13}$ under various magnetic fields parallel to $\left\langle 110 \right\rangle$, $\left\langle 111 \right\rangle$, and $\left\langle 100 \right\rangle$ axes.
All the results were obtained from the same crystal as the specific heat.
The anisotropy in the magnetization values was due to the effect of the crystal field. A transition is consistently observed up to 7 T.
The transition temperature decreased with increasing field strength; however, it was isotropic to the magnetic field.
Furthermore, the value of magnetization decreases with decreasing temperature near the transition temperature.
For these two reasons, we conclude that the ordered phase is three-dimensional antiferromagnetic.

\begin{figure}
\begin{center}
\includegraphics[width=9cm,clip]{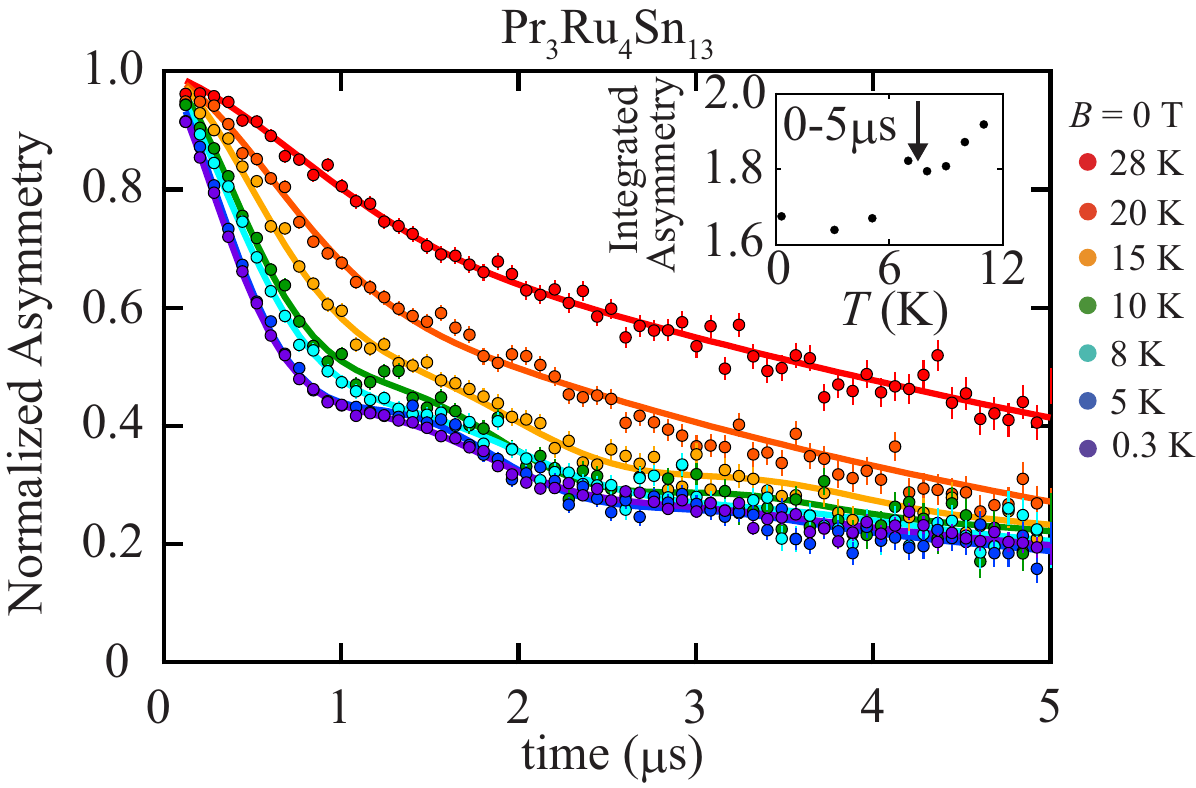}
\caption{\label{fig:crystal_structure} (Color online) $\mu$SR time spectra of Pr$_3$Ru$_4$Sn$_{13}$ under zero field. The solid curves are fitted results using Eq. (2) and the data collected above 0.3 K. The inset shows the temperature dependence of integrated asymmetry with time in the range $0-5$ {\textmu}s.
}
\end{center} 
\end{figure}

\subsection{$\mu$SR time-spectral analysis}
To  determine the reason for the small jump in specific heat at $T_N=7.5$ K, we discuss the results of $\mu$SR experiments under zero field.
To increase the amount of sample and obtain a strong signal, many small single crystals were crushed into a powder.
Figure 3 shows the temperature dependence of the $\mu$SR time spectra.
Below 20 K, the spectra were of the exponential type, suggesting the presence of an electron-fast depolarization component.
The inset shows the temperature dependence of integrated asymmetry with time in the range of $0-5$ {\textmu}s.
Around $T_N$, the integrals represented minimum.
Therefore, the phase is magnetically ordered at low temperatures.
Furthermore, there was faster longitudinal relaxation at lower temperatures, down to 0.3 K.
This indicates that the spin fluctuations coexist in a magnetically ordered state at low temperatures.

\begin{figure}[h]
\begin{center}
\includegraphics[width=9cm,clip]{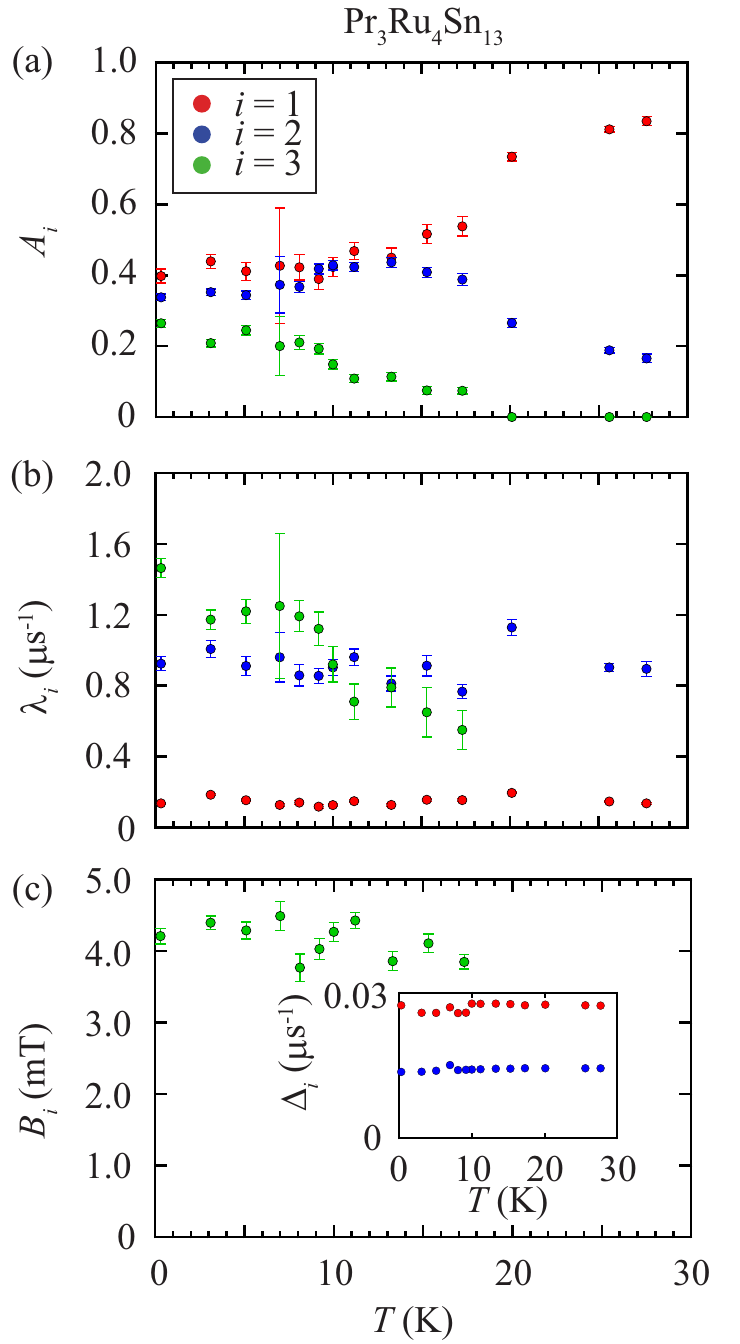}
\caption{\label{fig:crystal_structure} (Color online) Temperature dependences of (a) initial asymmetries, (b) muon-spin depolarization rates, and (c) inner fields estimated using Eq. (2) for Pr$_3$Ru$_4$Sn$_{13}$. The inset shows the temperature dependence of $\Delta _i$. Red, blue, and green circles represent $i = 1, 2,$ and 3.
}
\end{center} 
\end{figure}

We performed density functional theory (DFT) calculations using Winmostar program~\cite{win_2019} and found that there are two sites [(0, 0, 1/2) and (1/8, 1/8, 1/8)] that exhibit potential minima in the unit cell. These sites are at the center of Pr-Pr and Ru-Sn(1) bonds. Therefore, we performed our analysis using a function that assumes two muon-stopping sites.

\begin{equation}
\begin{split}
A(t) & =A_{1}e^{-\lambda_{1}t}G(\Delta _1,t)+A_{2}e^{-\lambda^2_{2}t^2}G(\Delta _2,t)+A_{3}\cos\left(\gamma B_{3}t \right)e^{-\lambda_{3}t},
\end{split}
\end{equation}
where
\begin{equation}
A_{1}+A_{2}+A_{3}=1,
\end{equation}
\begin{equation}
G(\Delta _i,t)=\frac{1}{3}+\frac{2}{3}\left(1-\Delta _i^{2}t^2\right)e^{-\frac{1}{2}\left(1-\Delta _i^{2}t^2\right)}.
\end{equation}
The first term is the fast relaxation of the nonmagnetic component due to nuclear spins at the one muon-stopping site. The second term is the slow relaxation of the depolarization component at the other muon-stopping site. The third term is the magnetic order component. When the volume fraction of magnetic order is 100\%, $A_3 = 2/3$ because the positrons counters are located at 0$^\circ$ and 180$^\circ$ to the muon beam. $G(\Delta _i,t)$ is a Kubo$\textendash$Toyabe function. $\gamma$ is the magnetic rotation ratio, $2\pi \times135.5$ MHz/T. To reproduce the shape of the spectrum, the first and second terms must be exponential and gaussian functions. Since the rotation term cannot be separated into two separate terms to analyze the experimental results, only the third term was used for the rotation term.

The inset of Fig. 4(c) shows the temperature dependence of $\Delta_i$. Reproducing time spectra for all temperatures and the fact that $\Delta_1$ and $\Delta_2$ are different values supports that there are two muon-stopping sites. Figure 4(a)-4(c) show the temperature dependence of $A_i$, $\lambda_i$, and $B_i$. Despite being above the transition temperature ($T_N = 7.5$ K), $A_3$ and $B_3$ are finite values below $\sim 17$ K. When the localized 4$f$ electrons of Pr are strongly coupled with the nucleus, an internal magnetic field is generated  at the muon-stopping sites, and an oscillation is also observed in the paramagnetic phase. This phenomenon has also been reported for PrNi$_5$~\cite{Kayzel_1994}, PrPb$_3$~\cite{Ito_2009}, and PrIr$_2$Zn$_{20}$~\cite{Higemoto_2012}. 
$A_3$ moves continuously between 10 K and 8 K. 
This change is due to spin relaxation originating from the magnetic phase transition. 
The volume fraction of the magnetic phase is $\sim 15$\% because the amount of change of $A_3$ between 10 K and 8 K is about $\sim 0.1$ with $A_3=2/3$ for 100\% volume fraction. 
This value is consistent with the result of the specific heat.
Therefore, the origin of the anomaly in specific heat is partial order transition.
This also supports the fact that $\lambda_i$ does not have a maximum at the transition temperature. Also, below 7 K, there was no clear decrease in lambda. If the electronic state is a singlet, the lambda is reduced. Thus, it suggests that spin fluctuations exist down to low temperatures.
We conclude that some of the 4$f$ spins are ordered and most of the 4$f$ spins are fluctuated.

\subsection{Chemical disorder effect}
As previously mentioned, the elemental ratio of our sample was Pr:Ru:Sn = 3:4:13.
Furthermore, the effective magnetic moment confirmed that the Pr ratio is 3.
In previous studies, the lattice constants of PrRuSn$_3$, which adopts the same crystal structure, were reported to be 9.709 \AA$ $~\cite{Eisenmann_1986} and  9.723 \AA~\cite{Fukuhara_1992}, which are similar to our results.
The structural difference between Pr$_3$Ru$_4$Sn$_{13}$ and PrRuSn$_3$ lies in the element occupying site 2a; PrRuSn$_3$ has a Pr-ion occupying site 2a.
This was also proposed for Pr$_3$Ru$_4$Ge$_{13}$~\cite{Ramakrishnan_1996}.
Additional Pr-ion is expected to induce a partial order in a small part of the 2a site in the Pr$_3$Ru$_4$Sn$_{13}$ crystal.

However, a  volume fraction with around ten percent cannot explain the EDX, and the effective magnetic moment results solely by the contribution of Pr-ion at the 2a site.
Since Pr is crystallographically one site and the effective magnetic moment is close to the theoretical value, the partial order is from a small amount of the additional Pr-ion.
This suggests that the spin of the Pr at the surrounding 6c site is also ordered due to the Pr at the 2a site through RKKY interaction.

\subsection{Quantum criticality in Pr3-4-13 family}
No magnetic transition was observed in Pr$_3$Co$_4$Sn$_{13}$ and Pr$_3$Rh$_4$Sn$_{13}$ when cooled to 0.2 K, and their ground states are reported to be singlet~\cite{Oduchi_2007,Nair_2018}.. 
Therefore, to move to QCP, the contribution of the RKKY interaction must increase. 
Elemental substitution is an effective method because it can apply negative and positive chemical pressures on magnetic ions.
Because the lattice constants of Pr$_3$Co$_4$Sn$_{13}$~\cite{Skolozdra_1983} and Pr$_3$Rh$_4$Sn$_{13}$~\cite{Nair_2018,Miraglia_1986,Hodeau_1982} are 9.57 and 9.69 \AA, respectively, the largest lattice constant, 9.72 \AA~for Pr$_3$Ru$_4$Sn$_{13}$, has more negative pressure effects than those of the other materials.
This effect is expected to cause Pr$_3$Ru$_4$Sn$_{13}$ to approach closest to the QCP, exhibiting behavior that cannot be explained by the singlet.
Supporting this prediction is that Pr$_3$Ru$_4$Sn$_{13}$ shows the strongest RKKY interaction, since Weiss temperatures of Pr$_3$Co$_4$Sn$_{13}$~\cite{Oduchi_2007}, Pr$_3$Rh$_4$Sn$_{13}$,~\cite{Oduchi_2007} and Pr$_3$Ru$_4$Sn$_{13}$ are -7, -7, and -23 K.
These scenarios can explain our results, in which the volume fraction of the partial order was increased by RKKY interaction.
Therefore, the additional Pr at 2a site causes the partial order of the surrounding 4$f$ electrons through RKKY interactions, but most of the 4$f$ electrons are fluctuated.
From our discussions, we conclude that Pr$_3$Ru$_4$Sn$_{13}$ is closest to the antiferromagnetically quantum critical point of Pr$_3$$T_4$Sn$_{13}$ ($T= $ Co, Ru, and Rh) for three reasons: (i) most of the 4$f$ electrons are fluctuated, (ii) the absolute value of the Weiss temperature of Pr$_3$Ru$_4$Sn$_{13}$ is the largest among Pr$_3$$T_4$Sn$_{13}$, and (iii) a strong RKKY interaction exists from the mechanism of partial order.

Finally, we discuss the quantum criticality of Pr3-4-13 family.
In Ce$_3$Co$_4$Sn$_{13}$, the coexistence of CDW and antiferromagnetic spin fluctuations have been reported.
It is important to find similar phenomena in the same crystal structure to investigate the CDW contribution.
We point out the possibility that spin fluctuations and CDW may coexist in the Pr3-4-13 family.
Furthermore, the synthesis of Pr3-4-13 with a large lattice constant would bring the family closer to QCP and advance the study of quantum criticality.
Therefore, Pr$_3$Ru$_4$Sn$_{13}$ provides a platform for the study of exotic quantum criticality.

\section{Summary}
Single crystals of Pr$_3$Ru$_4$Sn$_{13}$ were prepared, and single-crystal X-ray diffraction, XRD, NPD, magnetization, specific heat, and $\mu$SR measurements were performed.
The specific heat and magnetization exhibited an anomaly at $T_N = 7.5$ K.
From $\mu$SR time spectra, apparent relaxation and oscillation were observed.
However, the volume fraction of the magnetic phase estimated using the asymmetric value is around ten percent.
In addition, spin fluctuations were observed at low temperatures.
This provides microscopic evidence that the material is closest to the antiferromagnetically quantum critical point with a partial order of Pr$_3$$T_4$Sn$_{13}$ ($T= $ Co, Ru, and Rh).

\begin{acknowledgment}
We thank I. Watanabe, R. Kadono, T. U. Ito, Y. Ikeda, and Y. Nambu for the stimulating discussions, and M. Ohkawara for his technical support at AKANE. The neutron diffraction experiment at JRR-3 was conducted under a general user program managed by the Institute for Solid State Physics, University of Tokyo (Proposal Nos. 22409 and 23409) and supported by the Center of Neutron Science for Advanced Materials, Institute for Materials Research, Tohoku University. The $\mu$SR measurements were performed at the Materials and Life Science Experimental Facility of J-PARC (Proposal No. 2022B0290) using a user program. Magnetic susceptibility and heat capacity measurements were performed using PPMS and MPMS, respectively, at the CROSS User Laboratory and Institute for Materials Research, Tohoku University.
This work was financially supported by the JSPS/MEXT Grants-in-Aid for Scientific Research (Grant Nos. 21K13870 and 23K13051).
\end{acknowledgment}





\bibliography{Pr3Ru4Sn13}

\end{document}